\documentclass{aa}
\usepackage{pdflscape}
\usepackage{rotating}
\usepackage{longtable}
\usepackage{subfigure}
\usepackage{multirow}
\usepackage{color}
\usepackage{colortbl}
\usepackage{ctable}
\usepackage{xspace}
\usepackage[pdftitle={M.Cardillo-W44-V1}, colorlinks=true, citecolor=blue, linkcolor=blue, urlcolor=blue, pdfauthor={Martina Cardillo}]{hyperref}

\usepackage[varg]{txfonts}
\begin{document}

%\bibliographystyle{apj}
%\submitted{To be sumbmitted to \textit{A$\&$A}. Draft n.01 \today}

\title{The supernova remnant W44: a case of Cosmic-Ray reacceleration.}

\author{M.~Cardillo\inst{1}
\and E.~Amato\inst{1} \and P.~Blasi\inst{1,2}}% \and G.~Castelletti\inst{3} \and G.~Dubner\inst{3}}

\institute{INAF/Osservatorio Astrofisico di Arcetri, Largo E. Fermi, 5 - 50125 Firenze, Italy\\
e-mail:martina@arcetri.astro.it
\and Gran Sasso Science Institute (INFN), viale F. Crispi 7, 67100 L' Aquila, Italy.}%\\
%\and Instituto de Astronom\'ia y F\'isica del Espacio (IAFE), CC.67, Suc.28, 1428, Buenos Aires, Argentina.}

\date{Received / Accepted }

%****************************************abstract
\abstract{Supernova remnants (SNRs) are thought to be the primary sources of Galactic Cosmic Rays (CRs). In the last few years, the wealth of $\gamma$-ray data collected by GeV and TeV instruments has provided important information about particle energisation in these astrophysical sources, allowing us to make progress in assessing their role as CR accelerators. In particular, the spectrum of the $\gamma$-ray emission detected by AGILE and Fermi-LAT from the two middle aged Supernova Remnants (SNRs) W44 and IC443, has been proposed as a proof of CR acceleration in SNRs. Here we discuss the possibility that the radio and $\gamma$-ray spectra from W44 may be explained in terms of re-acceleration and compression of Galactic CRs. The recent measurement of the interstellar CR flux by Voyager I has been instrumental for our work, in that the result of the reprocessing of CRs by the shock in W44 depends on the CR spectrum at energies that are precluded to terrestrial measurement due to solar modulation. We introduce both CR protons and helium nuclei in our calculations, and secondary electrons produced {\it in situ} are compared with the flux of Galactic CR electrons reprocessed by the slow shock of this SNR. We find that the multi-wavelength spectrum of W44 can be explained by reaccelerated particles with no need of imposing any break on their distribution, but just a high energy cut-off at the maximum energy the accelerator can provide. We also find that a model including both re-acceleration and a very small fraction of freshly accelerated particles may be more satisfactory on physical grounds.}

\keywords{Cosmic-rays, Acceleration of particles, Supernova remnants, Molecular processes, Shock waves, Radiation mechanisms: non-thermal}

\titlerunning{The SNR W44}
\authorrunning{M.Cardillo}

\maketitle

%%%%%%%%%%%%%%%%%%%%%%%%%%%%%%%%%%%%%%%
\section{Introduction}
\label{Introduction}
%%%%%%%%%%%%%%%%%%%%%%%%%%%%%%%%%%%%%%% 
\label{sec:intro}

%---------------------------
SNR W44 is a middle-aged, [$10,000\div 20000$] yrs old \citep{smith85_W44,wolszczan91_W44}, SNR, located in the Galactic Plane ($l,b$)=~($34.7,-0.4)$ at a distance $d\sim2.9$~kpc from Earth \citep[][ and references therein]{castelletti07}. This source has been studied at all wavelengths: it is characterized by a quasi-elliptical shell in the radio band \citep{castelletti07} and it is expanding with a velocity $v_{sh}\sim 100\div150$ km$/$s \citep{reach00_W44} in a molecular cloud (MC) complex with an average proton density of $n\sim 200$ cm$^{-3}$ \citep{wootten77,rho94,yoshiike13_W44}.
The $\gamma$-ray emission detected from the South-East part of this remnant \citep{giuliani11_W44, abdo10_W44, ackermann13_W44, cardillo14_W44}, a region that has an embedded MC, is likely connected with the interaction between the SNR shock and the cloud, highlighted by the presence of OH maser (1720~MHz) emission \citep{claussen97,hoffman05}.
W44 became the center of attention of the scientific community when \cite{giuliani11_W44} published its detection with the AGILE $\gamma$-ray satellite at photon energies below 200 MeV, making it the first SNR ever detected in this energy range. These observations have been especially important because the emission falls in an energy interval where it is actually possible to distinguish the $\gamma$-ray contribution due to leptonic emission processes (Bremsstrahlung or IC) from the one of hadronic origin, due to the decay of neutral pions originating in CR interactions with ambient matter. Indeed the $\gamma$-ray spectrum of W44 showed the pion bump that can be unequivocally linked to CR hadronic interactions, as later confirmed by the Fermi-LAT measurements \citep{ackermann13_W44}. The same Fermi-LAT paper claimed detection of another middle-aged SNR, IC443, where the pion bump was also found. Additional AGILE observations of W44 \cite[]{cardillo14_W44} showed that only models in which the $\gamma$-ray emission is of hadronic origin are able to explain the multi wavelength spectrum of this remnant from radio \citep{castelletti07} to $\gamma$-rays.

In fact, all these models, while reproducing the multi-wavelength spectrum of W44, have some puzzling features, in that the fit to experimental data requires accelerated protons with a broken power-law spectrum at low energies and a very steep high energy spectral index.

The possible reacceleration and compression of pre-existing CRs was first considered by \cite{uchiyama10_W44} and \cite{lee15_W44} who showed that the shock propagating into the cloud crushed by the remnant \citep{blandford82} may explain the spectral properties of SNR~W44. In these calculations the Galactic CR spectrum was parametrized as in \cite{strong04} and \cite{shikaze07}, with the addition of a low energy cutoff and a high energy steepening of the reaccelerated spectrum. The latter was modeled following the idea of the so called Alfv\'en wave damping put forward by \cite{malkov11_W44}, having in mind the very steep gamma ray spectrum inferred from AGILE and Fermi-LAT data.

Very recently, a time dependent model of reacceleration was developed by \cite{tang15}. The authors follow the time evolution of the reaccelerated particle spectrum, considering different energy dependencies of the diffusion coefficient. The leptonic contribution to the emission, coming in principle both from primary and secondary electrons, and showing signatures both in the radio and in the gamma-ray band, is not taken into account. The gamma-ray emission is explained as a result of compression and reacceleration of galactic CR protons, without the need of any additional break, such as that coming from Alfv\`en wave damping, but particle energy losses are not taken into account either.

In the present paper we developed further the idea that the multi-wavelength spectrum of W44 can be explained as the result of re-acceleration, improving on existing models in several respects: 1) we use actual data on the low energy CR spectrum in the local interstellar medium (LIS) as provided by Voyager I  \citep[]{potgieter14}, instead of assuming a parametrization of the demodulated CR spectrum. 2) We use both spectra of protons and helium nuclei (from Voyager I) for the production of gamma rays from pion decays, as well as for the production of secondary electrons and positrons. The flux of such secondary leptons is then compared with the spectrum of reprocessed Galactic CR electrons (again inferred, though more indirectly, from Voyager data). 3) We use radio data that are limited to the same spatial region where the AGILE and Fermi-LAT emission appears to be originating, thereby having direct control upon the filling factor of the emitting region. 4) We test the possibility that the spectral steepening that both AGILE and Fermi-LAT infer from their data at high energy reflects a cutoff in the spectrum of re-energized Galactic CRs, rather than some sort of plasma effect, as postulated by \cite{malkov11_W44}. Such a spectral feature naturally arises from the overlap of reacceleration (up to a maximum energy of $\sim 10$ GeV) and compression of Galactic CRs in the crushed cloud.

Here we confirm that the radio and gamma ray data from W44 can be explained in terms of reacceleration and compression of galactic CRs, provided the maximum energy of reaccelerated particles is of order $\sim 10$ GeV. This explanation requires however that a relatively large fraction of the surface of the shock, $\sim 50\%$, is interacting with the dense cloud. We investigated whether this requirement could be alleviated by accounting for the likely inhomogeneity of the cloud, which is expected to be made of very dense clumps, with densities reaching $\sim 2000\ {\rm cm}^{-3}$ and a more diluted interclump medium with density $\sim 20\ {\rm cm}^{-3}$. We found that such a density structure leads to estimate even larger filling factors to explain the gamma-ray emission, while the problem could be easily alleviated by assuming that, in addition to the dominant reaccelerated component, a very small fraction, of order $10^{-4}$, of the shock kinetic energy is converted into acceleration of fresh particles.  

Finally, given the possibility that the reacceleration be weak in the slow shock of W44 and that even a small fraction of neutrals in the pre-shock medium could damp the magnetic perturbations necessary for any acceleration (or reacceleration) process to take place, we analyse the case of pure compression of Galactic CRs in the crushed cloud. We find that with a density of order $\sim 10^{4}cm^{-3}$ and a magnetic field strength $\sim 1$ mG a decent fit to the data can be found even in this case.

The paper is organised as follows: in \S~\ref{Sec:Clouds} we describe the essential physical aspects of the crushed-cloud scenario; in \S\ref{Sec:REvsA} we derive the reaccelerated and accelerated particle spectra after compression and energy losses; the calculation of the equilibrium particle distributions in the crushed cloud is illustrated in \S \ref{sec:Losses}. A complete analysis of our results is shown in  \S~\ref{Sec:Results}. Our conclusions are drawn in \S \ref{sec:Conclusions}.

%%%%%%%%%%%%%%%%%%%%%%%%%%%%%%%%%%%%%%%
\section{Middle aged SNRs interacting with molecular clouds}  
\label{Sec:Clouds}
%%%%%%%%%%%%%%%%%%%%%%%%%%%%%%%%%%%%%%%
A number of SNRs detected so far in the $\gamma$-ray band are middle-aged sources, bright in the radio band as well, and they are all interacting with molecular clouds \citep[e.g.,][]{giuliani10_W28,aleksic12_W51,ackermann13_W44,cardillo14_W44}. The pioneering work of \cite{blandford82} on clouds crushed by a supernova explosion showed that re-energization of pre-existing CRs in the interaction region can account for the radio emission. Asking whether similar processes may also explain the gamma ray emission appears only natural. 

Here we briefly summarize the picture that \cite{blandford82} proposed to describe the interaction between a SNR shock and a molecular cloud: if the density of the pre-shock material and the shock velocity are large enough, then the material behind the shock is compressed enough to make it radiative and the resulting radiation can ionise the whole region where the shock is propagating. Immediately behind the shock the gas is simply compressed. When recombination starts and the region becomes radiative, the gas becomes more compressive and its density increases further: for a shock velocity $v_{sh7}\simeq100$ km/s, as inferred for W44, the critical column density for the formation of a radiative region is $N_{cool}\simeq3\times10^{17}v_{sh7}^{4}$ cm$^{-2}$ \citep{mckee87}. This condition translates into a lower limit for the density of the cloud:
%**********
\begin{equation}
n_{cloud}\gtrsim n_{cool}\equiv10\,E_{51}^{0.81}\bar{n}_{-1}^{-0.19}R_{1}^{-2.8}\,\,\,cm^{-3}
\label{Eq:com_limit}
\end{equation}
%****************
where $E_{51}$ is the SNR energy in units of $10^{51}$ erg, $\bar{n}_{-1}$ is the effective mean hydrogen density in the remnant in units of $0.1$ cm$^{-3}$ and $R_{1}$ is the SNR shock radius in units of $10$ pc \citep{blandford82}. The compression factor between downstream of the shock (density $n_{d}$) and the crushed cloud (density $n_{m}$), $s\equiv \frac{n_{m}}{n_{d}}$, can be limited by magnetic or thermal pressure, depending on the value of the pre-shock magnetic field and cloud density. The compression associated to radiative cooling causes energization of pre-existing particles, boosting the normalisation of the particle spectrum by a factor $s^{2/3}$ (pitch angle isotropization is assumed), while the momentum per particle increases as $p\rightarrow ps^{1/3}$.

%%%%%%%%%%%%%%%%%%%%%%%%%%%%%%%%%%%%%%%
\section{Reacceleration, acceleration and compression}  
\label{Sec:REvsA}
%%%%%%%%%%%%%%%%%%%%%%%%%%%%%%%%%%%%%%%

%-------------------------------
\subsection{Reacceleration of Galactic CRs}
\label{Sec:Reacc}
%--------------------------------------------
CRs permeating the cloud are reaccelerated at the SNR shock independent of their energy, since they are already non-thermal, and there should be no obstacle to inject them at the shock. Even slow shocks are expected to contribute substantial re-acceleration, while they are not expected to be efficient at accelerating fresh particles.
 
Here we compute the reaccelerated spectra downstream of the shock by using, for convenience, the formalism introduced by \cite{Blasi04}, where however the more complex case of non-linear shock reacceleration was considered. We solve the stationary transport equation
%*****************
\begin{equation}
u\frac{\partial f}{\partial x}=\frac{\partial}{\partial x}\left[D(p)\frac{\partial f}{\partial x}\right]+\frac{1}{3}\left(\frac{\partial u}{\partial x}\right)p\frac{\partial f}{\partial p},
\label{eq:transport}
\end{equation}
where $f(x,p)$ is the CR distribution function, $u$ is shock velocity and $D(p)$ the energy dependent diffusion coefficient. The $x$ axis is parallel to the shock normal and the shock is located at $x=0$. The boundary condition at upstream infinity (located at $x=-\infty$) is imposed by requiring that the distribution function, $f_{\infty}(p)$, equals the Galactic CR distribution, namely $f(x=-\infty,p)=f_{\infty}(p)$.

The solution of the transport equation at the shock location, $f_{0}(p)=f(x=0,p)$, is easily found in the form:
\begin{equation}
f_{0}(p)=\alpha\left(\frac{p}{p_{m}}\right)^{-\alpha}\int^{p}_{p_{m}}\frac{dp'}{p'}\left(\frac{p'}{p_{m}}\right)^{\alpha}f_{\infty}(p')
\label{eq:Re}
\end{equation}
where $\alpha= \frac{3v_{sh}}{v_{sh}-u_{d}}$ and $u_{d}$ is the gas velocity downstream of the shock. The momentum $p_{m}$ in Eq. \ref{eq:Re} represents a minimum momentum in the spectrum of Galactic CRs. For our purposes, the value of $p_{m}$ is so low that it does not have any effect on the results that we present below. 

One can see that the effect of reacceleration is twofold: first, it increases the momentum per particle up to a maximum momentum that is determined by the balance between acceleration time and age of the system. Second, it hardens the spectrum of the reaccelerated particles, provided the spectrum of Galactic CRs is steeper than $p^{-\alpha}$. If this latter condition is not satisfied, then the reaccelerated spectrum has the same slope as the original spectrum, but a different normalisation.

We adopt a parametrisation of the Galactic CR spectrum that reflects the recent measurements from the Voyager I at low energies ($E\gtrsim 1$ MeV/n)\citep[]{webber13_Voyager} and the PAMELA \citep{Adriani11_p} and AMS-02 data \citep{Aguilar15_e,Aguilar15_p,Aguilar15_He} at higher energies. The Voyager I measurements carried out after 2013, namely after the spacecraft crossed the boundary of the heliosheath (HS) and entered the heliopause (HP), provides us with the first direct knowledge of the spectrum of Galactic CR protons and helium nuclei in the LIS and makes it possible to avoid calculating the effects of solar modulation. Some corrections to take into account solar modulation are instead needed to infer the electron spectrum  \citep{potgieter13_e}, since Voyager data on electrons date back to 2010, when it was at $~112$ AU from Earth.

\begin{figure}[ht!!!!!]
\centering
\includegraphics[scale=0.85]{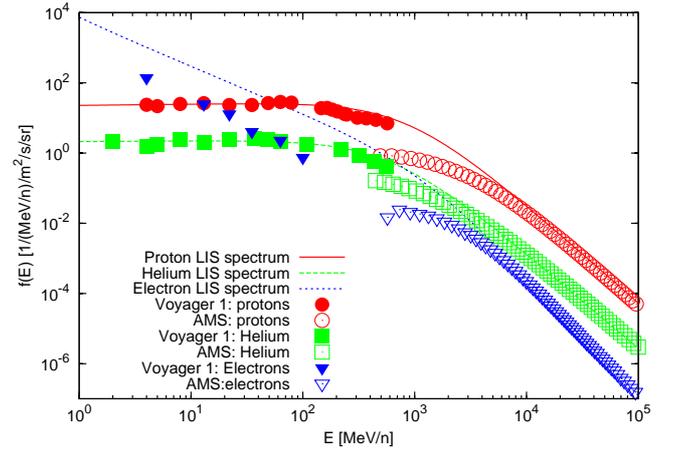} 
\caption{Protons (red), Helium (green) and electrons (blue) spectra. Circles represent protons, squares electrons and triangles Helium data points; measurements from Voyager~1 \citep{potgieter14} are represented by filled points, whereas AMS measurements \citep{Aguilar15_e,Aguilar15_p,Aguilar15_He} are represented by open points. The curves represent the fits to the LIS spectrum of the different species used in this work\citep{potgieter14}. The blue curve does not go through the Voyager~1 data points because the electrons' data are those collected in 2010, when Voyager had not entered the HP yet. }
\label{Fig:Potgieter}
\end{figure}

The flux of electrons in the LIS is well described as
%***************
\begin{equation}
J_{LIS,e}= 0.21\left(\frac{E^{1.35}}{\beta_{e}^{2}}\right)\left(\frac{E^{1.65}+0.6920}{1.6920}\right)^{-1.1515}+J_{bump}
\label{Eq:LIS_e}
\end{equation}
%*********************
in units of $[{\rm particles}/{\rm m}^{2}/{\rm s}/{\rm sr}/{\rm MeV}]$ \citep{potgieter13_e} . Here $E$ is the electron kinetic energy expressed in units of GeV whereas $J_{bump}$ is introduced in order to fit PAMELA (and AMS02) data in the $[5-20]$ GeV range and has the following form:
%****************
\begin{equation}
J_{bump}=1.73\, \exp\left[4.19-5.40\, \ln\,(E)-8.9\,E^{-0.64}\right]\ .
\label{Eq:LIS_ebump}
\end{equation}
%******************
The spectrum of protons and He nuclei measured by Voyager~1 in 2013, after entering the HP, is well represented as \citep{bisschoff15}:
%*********************
\begin{equation}
J_{LIS,n}=A_h\, \left(\frac{E^{a}}{\beta_{p}^{2}}\right)\left(\frac{E^{d}+k^{d}}{1+k^{d}}\right)^{-b}\ ,
\label{Eq:LIS_p}
\end{equation}
%********************
with $J_{LIS,n}$ in units of $[particles/m^{2}/s/sr/(GeV/n)]$ and $E$ the particle kinetic energy per nucleon in units of $[GeV/n]$. The parameters appearing in Eq.~\ref{Eq:LIS_p} are: $a=1.02$, $b=3.15$, $d=1.19$, $k=0.60$, $A_h=3719$ for the proton LIS, $J_{LIS,p}$; and $a=1.03$, $b=3.18$, $d=1.21$, $k=0.77$, $A_h=195.4$ for the helium LIS, $J_{LIS,He}$.
In order to include all the features of the Galactic CR spectrum, we also impose the spectral hardening of the proton and helium spectra detected by PAMELA \citep{Adriani11_p} and confirmed by AMS \citep{Aguilar15_p,Aguilar15_He}. The latest data by AMS02 show a hardening by $\Delta\gamma_{p}=0.119$ for protons of rigidity larger than $R_{0,p}=336$ GV, and by $\Delta\gamma_{He}=0.133$ for helium nuclei at rigidity larger than $R_{0,He}=245$ GV. The final form of our galactic particle spectrum, for type $j$ nuclei, is:
%*********************
\begin{equation}
J'_{LIS,j}(E)=J_{LIS,j}\times\left[1+\left(\frac{E_{i}}{E_{0,i}}\right)^{\frac{\Delta\gamma_{i}}{s}}\right]^{\sigma}
\label{Eq:LIS_hard}
\end{equation}
%***********************
where $E_{0,i}$ is the kinetic energy that corresponds to a rigidity $R_{0,i}$ and $\sigma$ is a smoothing parameter that we take as $\sigma=0.024$.
The spectra of particles of all species $i$ that we obtain from $J_{LIS,i}$ are shown in Fig.~\ref{Fig:Potgieter}: protons are in red, helium nuclei in green and electrons in blue; overplotted are also the Voyager~1 and AMS02 data, as triangles and circles respectively. The reason why the electron curve appears to not fit the data is that the curve represents the very LIS spectrum while data are those collected by Voyager~1 in 2010, before entering the HP. For protons and helium, instead, we could use the HP spectra \citep{potgieter14}. The CR spectra $f_{\infty,i}(p)$ to be used in our calculations are related to the fluxes $J_{LIS,i}(E)$ by the usual expression: 
\begin{equation}
4 \pi p^{2} f_{\infty,i} (p) dp = \frac{4\pi}{v(p)} J_{LIS,i} (E) dE,
\end{equation}
where $v(p)$ is the particle velocity at momentum $p$.

%------------------------ Figure injection spectra reacceleration
%-------------------------------------------

%-------------------------------------------
\subsection{Acceleration}
\label{Sec:Acc}
%------------------------------
The other possible contribution to the $\gamma$-ray emission of SNR W44, and the one that was actually considered first, comes from fresh acceleration of particles at the shock. In the most general case, a simple power-law distribution is assumed for both the hadronic and leptonic components:
%******************
\begin{equation}
f_{i}(p)=k_{i}\left(\frac{p}{p_{inj}}\right)^{-\alpha}
\label{eq_acc_distribution}
\end{equation}
%******************
where the index $i$ reads $p$ for protons and $e$ for electrons, $\alpha$ is the spectral index, that depends on the shock compression ratio and is the same for all species, and $p_{inj}$ is the injection momentum corresponding to an injection energy $E_{inj}\sim4.5E_{sh}$, with $E_{sh}=\frac{1}{2}m_{p}v_{sh}^{2}$ \citep{caprioli14_I}. Finally, $k_{i}$ is a normalization constant related to the CR acceleration efficiency, $\xi_{CR}$. We define the latter as the fraction of shock ram pressure, $\rho_0 v_{sh}^2$, that is converted into CR pressure, $P_{CR}$:
%******************
\begin{equation}
P_{CR}=\frac{4\pi}{3}\int f_{p}(p)p^{2}pv(p)dp=\xi_{CR}\rho_{0}v_{sh}^{2}\ ,
\label{eq:pressure}
\end{equation}
%******************
where $\rho_{0}$ is the upstream density and $v_{sh}$ the shock velocity. Substituting the proton spectrum of Eq.~\ref{eq_acc_distribution} in Eq.~\ref{eq:pressure}, we obtain the relation between $k_p$ and $\xi_{CR}$:
%******************
\begin{equation}
k_{p}=\frac{\xi_{CR}\rho_{0}v_{sh}^{2}}{\frac{4\pi}{3}\int \left(p/p_{inj}\right)^{-\alpha}p^{3}v(p)dp}\ .
\end{equation}
%******************
The normalization of the electron distribution, $k_{e}$, is then fixed by assuming the standard CR electron/proton ratio, $k_{ep}\approx10^{-2}$, and that the electron spectrum follows that of protons.

\subsection{Compression}
\label{Sec:Comp}
%------------------------------
As anticipated in Sec.~\ref{Sec:Clouds}, both freshly accelerated and reaccelerated particles experience further energisation in the recombination region due to compression. The effect is described by the factor $s$:
%******************
\begin{equation}
s\equiv \left(\frac{n_{m}}{n_{d}}\right)=\left(\frac{n_{m}}{r_{sh} n_{0}}\right)
\label{Eq:factor s}
\end{equation}
%******************
where $n_{d}$ is the density immediately downstream of the shock, $n_{m}/n_d$ is the compression due to radiative cooling and $r_{sh}=n_{d}/n_{0}$ is the compression ratio at the shock. Following the considerations put forward by \cite{blandford82}, we assume that in the SNR~W44 the gas compression due to cooling is limited by magnetic pressure (see Sec.~\ref{Sec:Clouds}) and we compute the value of $n_m$ by equating the magnetic pressure to the shock ram pressure \citep{uchiyama10_W44}:
%******************
\begin{equation}
\frac{B_{m}^{2}}{8\pi}=n_{0}\mu_{H}v_{sh}^{2}\rightarrow n_{m}\simeq94\left(\frac{n_{0}}{1\,cm^{-3}}\right)^{3/2}\left(\frac{B_{0}}{1\,\mu G}\right)^{-1}\left(\frac{v_{sh}}{10^{7}cm/s}\right),
\label{Eq:nm}
\end{equation}
%******************
where $B_{m}=\sqrt{\frac{2}{3}}\left(\frac{n_{m}}{n_{0}}\right)B_{0}$ is the compressed magnetic field, $\mu_{H}$ is the mass per hydrogen nucleus and $B_{0}=b\sqrt{n_{0}/cm^{-3}}$ $\mu$G is the unperturbed magnetic field upstream of the shock. The term $b=V_{A}/(1.84\,km/s)$ is of order $\sim 1$ in the ISM and it is in the range [0.3-3] in MCs, depending on the Alfv\' en velocity $V_{A}$ \citep{hollenbach89,crutcher99}.

{ As a result of compression, the spectrum of particles changes according to
%Compression affects both the momentum and the spectrum of particles. Indeed, reaccelerated and accelerated particles gain momentum because of the higher density, $p\,\rightarrow s^{1/3}p$, so that their final spectrum becomes:
%******************
\begin{equation}
f'(p)=f_{0}(s^{-1/3}p)\ ,
\label{Eq:compressed spectrum}
\end{equation}
%******************
where $f_{0}(p)$ is the spectrum of CRs resulting from reacceleration and shock compression. In other words, $f_{0}$ has two components: one given by Eq. \ref{eq:Re} with an exponential cutoff at a maximum momentum $p_{M}$ that will be estimated in \S \ref{Sec:Results}, and another component that simply results from compression of Galactic CRs at the shock. As an alternative one can directly account for the compression between upstream and the crushed cloud by considering $s'=n_{m}/n_{0}$ and calculating the compressed component as $f_{comp}(p)=f_{\infty}(s'^{-1/3}p)$, for $p\gg p_{M}$. 

%-------------------------------------------
\section{Equilibrium particle distribution in the crashed cloud}
\label{sec:Losses}
%------------------------------
In order to estimate the $\gamma$-ray flux produced by reaccelerated and accelerated CRs, we need to take into account the energy losses that affect protons,  helium nuclei and electrons in the environment that we are considering.

High energy protons mainly lose energy because of pp-interactions (at high energies) and ionization (at lower energies). In particular, ionization losses are important for all species, protons, helium nuclei and electrons. Since in SNR~W44 the shock velocity is $v_{sh}\geq100-150$ $km/s$, we assume that the gas around the shock is completely ionized \citep{blandford82}. Ionisation losses are described as discussed in detail in Appendix \S \ref{Sec:app1} (Eqs.~\ref{eq:p_ion} and \ref{eq:e_ion}). The effect of ionization on the particle spectrum is that of leading to a low energy hardening. Electrons are also affected by synchrotron, Bremsstrahlung and IC losses (see Eqs.~\ref{eq:sync_loss}, \ref{eq:ic_loss}, \ref{eq:Brems_loss_low} in Appendix \ref{Sec:app1}).

In order to obtain the final particle spectrum, we need to solve the kinetic equation for every particle of specie "i":
%****************
\begin{equation}
\frac{\partial N_{i}(E,t)}{\partial t}=\frac{\partial}{\partial E}\left[b(E)N_{i}(E,t)\right]+Q_{i}(E),
\label{Eq:kinetic}
\end{equation}
%****************
where $b(E)=-\frac{dE}{dt}$ is the sum of all the loss terms listed above, and $Q_{i}(E)$ is the particle injection rate per unit energy interval.
The particle spectra are obtained by solving the kinetic equation numerically for light nuclei, primary electrons and secondary electrons (produced from pion decay following p-p scattering).

For hadrons and primary leptons, the injection term reads: 
%****************
\begin{equation}
Q_{i}(E)=\left(\frac{n_{0}}{n_{m}}\right)\frac{1}{t_{int}}N'_{i}(E),
\label{eq:injection_rate}
\end{equation}
%****************
with $N'_{i}(E)=4 \pi p^2 f'(p) dp/dE$ the reaccelerated and compressed spectrum (obtained from $f'(p)$ in Eq.~\ref{Eq:compressed spectrum} after transforming to energy space). The term $t_{int}$ is the interaction time between the cloud and the remnant that we assume to be less than the SNR age; indeed, even if the source is middle-aged ($t_{age}\sim1.5\times 10^{4}$ yrs), the interaction has likely started more recently.

As for secondary electrons, their injection derives from the decay of charged pions produced in inelastic nuclear collisions of energetic hadrons. The injection rate to be used in Eq.~\ref{Eq:kinetic} can be written in this case as 
\begin{equation}
Q_s(E)=
\left\{
\begin{array}{ll}
\Phi_{e,low}, & E<0.1 TeV\\
 & \\
\Phi_{e,high},& E>0.1 TeV
\end{array}
\right.
\label{eq:qsec}
\end{equation}
where $\Phi_{e,low}$ and $\Phi_{e,high}$ are given in Eqs.~\ref{Eq:sec_low} and \ref{Eq:sec_high}, respectively.}

%%%%%%%%%%%%%%%%%%%%%%%%%%%%%%%%%%%%%%%%
\section{Results}
\label{Sec:Results}
%%%%%%%%%%%%%%%%%%%%%%%%%%%%%%%%%%%%%%%

In this section we illustrate our results in terms of reacceleration and additional compression of Galactic CRs in the crashed cloud of SNR W44, and compare them with the gamma ray emission as measured by AGILE \cite[]{cardillo14_W44} and by Fermi-LAT \cite[]{ackermann13_W44}, and with the radio emission as measured by VLA \cite[]{castelletti07} and Planck \citep[]{planck14}. Below, when we refer to protons, helium nuclei and electrons we actually refer to Galactic CR particles that have been reaccelerated at the W44 shock and further compressed in the crashed cloud. The role of secondary electrons and positrons is singled out and commented upon whenever appropriate. 

As far as gamma ray emission is concerned, we include the contribution due to pion production (from both protons and helium nuclei) and bremsstrahlung emission of both electrons and secondary $e^{+}-e^{-}$ pairs produced in inelastic hadronic interactions.

In order to compare emission from the same accelerated particles, we limited the fit to the radio emitting volume as defined by the gamma ray emission region. The radio data limited to such area have been provided to us by Dr. Castelletti and Dr. Dubner through a private communication. 

In our calculation, we kept fixed all the SNR parameters for which we have reliable estimates: its distance, $d=2.9$ kpc, and consequently its size in the radio waveband, $R_{SN}=12.5$ pc, \citep{castelletti07}, the average gas density, $n_{0}\sim200$ cm$^{-3}$ \citep{yoshiike13_W44} and the shock velocity, $v_{sh}\sim100-150$ km/s \citep{reach00_W44}. In order to tune our predictions to the data, we use as free parameters the interaction time between the remnant and the cloud, $t_{int}$ ($\le t_{age}$), and the pre-shock magnetic field strength (parametrised through $b$, see Sec.~\ref{Sec:Comp}).

Moreover, the emissivity at all frequencies should be integrated over the emission volume. With respect to previous analyses, we parametrize the emission volume in a different way, and a way that, as we discuss in the following, brings to our attention a potential weakness of the reacceleration model. Since the emissivity is assumed to be uniform over the crushed cloud volume, we multiply the emissivity by the emission volume taken to be $4\pi\xi R_{SN}^{2} v_{sh} t_{int}$, where $\xi<1$ is the fraction of the spherical surface of the SN shell that is covered by the cloud. In previous work, the authors had introduced a filling factor $f<1$ defined as the fraction of the whole SNR volume, $\frac{4}{3}\pi R_{SN}^{3}$, occupied by the crushed cloud. The two parameters are related by the simple expression:
$$
f = 3~\xi ~\frac{v_{sh}t_{int}}{R_{SN}}.
$$
It is easy to see that for $R_{SN}=12.5$ pc, $v_{sh}\sim100$ km/s and $t_{int}=t_{age}/2\approx 5000$ yr, as used by \cite{uchiyama10_W44}, the value of $f\approx 0.2$ adopted by the same authors corresponds to $\xi\approx 12.5>1$. This discrepancy is due to the fact that, although $f<1$, the volume of the remnant could not possibly have been filled by the shock moving at $100$ km/s, so that in terms of surface of the SN shell filled by the crushed cloud, this value of $f$ corresponds to $\xi>1$. This implies that the values of the parameters to be assumed in the calculations are bound to be different from the ones chosen by  \cite{uchiyama10_W44} (a similar line of argument holds for the results of \cite{lee15_W44}). We will see that even in this case it is not easy to obtain a set of fitting parameters for which $\xi<0.5$, namely a large fraction of the surface is required to be covered by the crushed cloud.

It is straightforward to realize, by looking at Eq.~\ref{Eq:nm}, that, due to the assumption that radiative compression is limited by magnetic field pressure, the upstream magnetic field plays a crucial role in determining the compressed density $n_m$, which in turn affects ionisation losses and the rate of nuclear collisions. In addition, $B_0$ enters, together with $t_{int}$, the expression for the maximum energy of reaccelerated or accelerated particles.

Since energy losses are not very important for high energy protons and wave damping is also unimportant, if a fully ionized pre-shock medium is assumed, the high-energy cut-off of the accelerated particle spectrum is determined by the lifetime of the accelerator, namely by the condition that $t_{acc}(E_{max})<t_{int}$. The same is true for reaccelerated electrons, namely the maximum energy is not determined by balancing acceleration rate with radiative losses, but it is simply dominated by the finite acceleration time. The latter can be written, as a function of the shock velocity and particle diffusion coefficient, $D(p)$, as $t_{acc}\approx D(p)/v_{sh}^{2}$. We then write the diffusion coefficient using quasi-linear theory, assuming $\delta B/B_0\sim 1$ at $k_0=1/L_c$ with $L_c$ the coherence length of the field in the crashed cloud:
%%%%%%%%%%%%%%%
\begin{equation}
D(E)=\frac{1}{3}r_{L}c\left(\frac{L_c}{r_L}\right)^{\delta}\ .
\label{eq:de}
\end{equation}
%%%%%%%%%%%%%%%
In Eq.~\ref{eq:de} $r_L$ is the particle Larmor radius and $\delta = k_T-1$, with $k_T$ the turbulence spectral index. For a Kolmogorov perturbation spectrum, as we assume in our model, $k_T=5/3$ and the maximum momentum reads:
%%%%%%%%%%%
\begin{equation}
p_{\rm max}\sim7\,GeV/c\,\left(\frac{B_{0}}{30\,\mu G}\right)\left(\frac{v_{sh}}{130 \,km/s}\right)^{6}\left(\frac{t_{int}}{15000\,yrs}\right)^{3}\left(\frac{L_{c}}{0.1\,pc}\right)^{-2}\ ,
\label{Eq:EmaxKo}
\end{equation}
%%%%%%%%

The strongest dependence is on shock velocity, that is also the only quantity appearing in Eqs.~\ref{Eq:EmaxKo} for which we have a direct measurement from line emission \citep{reach00_W44}.

Free parameters are instead $B_0$, $L_c$ and $t_{int}$. The latter is constrained to be $t_{int}<t_{age}\approx 1.5\times10^4\ yr$, but in reality a stronger constraint comes from fitting the source SED, since the ratio of secondary-to-primary electrons increases with it. As to the pre-shock magnetic field, our best fit of the multi wavelength spectrum of the source is obtained for $B_0=34\ \mu$G. The corresponding post-shock compressed field reaches a value $B_{m}\approx 1.4$ mG and the post-shock compressed density is $n_m\approx 10100\ {\rm cm}^{-3}$, about 2 orders of magnitude larger than the pre-shock average density.

Concerning the correlation length, in the ISM this is typically taken to be $L_{c}=50-100$ pc, the scale at which turbulence is assumed to be injected by SNe. In molecular clouds, however, $L_c$ is thought to be much smaller: from measurements of dust polarization \citep[][and therein]{houde09}, we know that $L_{c}$ can reach values as small as $0.1-0.01$ pc.

Since the cut-off is at relatively low-energies, the compression ratio that can be used in order to fit the radio and $\gamma$-ray data can be in a range $r_{sh}=3.5\div4$ without changes in the fit, implying a momentum spectral index $\alpha=4.2\div4$.

%-------------------------------
\subsection{Reacceleration Model}
\label{ReaccMod}
%--------------------------------------------
%------------------------ Figure final spectra reacceleration
\begin{figure}[ht!!!!!]
\centering
\includegraphics[scale=0.85]{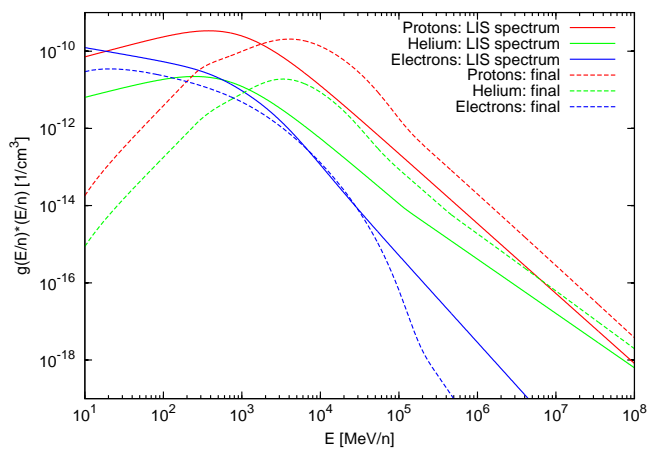} 
\caption{Spectra of protons, He nuclei and electrons in the local ISM (solid lines) and after shock reacceleration and compression in the crushed cloud (dashed lines). The change of slope at $\sim 10$ GeV is associated with the acceleration cutoff. Particles at higher energy are simply compressed in the cloud. Ionisation energy losses affect the spectral slopes at lower energies, synchrotron losses steepen the electron spectrum at higher energies, whereas Bremsstrahlung and pp losses mainly affect the normalization.}
\label{Fig:final_reacc_spectra}
\end{figure}
%-------------------------------------------
The diffuse CR protons, He nuclei and electrons in the ISM are reaccelerated at the shock of W44 up to a maximum energy discussed above. In addition, they are compressed adiabatically both at the shock and in the crushed cloud. The penetration of CRs in the dense cloud is also accompanied by energy losses that result in further spectral changes. The spectrum of all species resulting from all these processes is illustrated in Fig.~\ref{Fig:final_reacc_spectra} as dashed lines and should be compared with the ISM spectra (solid lines). The curves are obtained with a combination of parameters that, as we show below, provides the best fit to the SED of W44: $t_{int}=8400$ yr, $n_{0}=200~\rm cm^{-3}$, $B_{0}=34\mu G$ (corresponding to $b=2.4$), $v_{sh}=130$ km/s. With these values of the environmental parameters, one deduces a compressed density in the cloud of $\sim 10^{4}~\rm cm^{-3}$, a magnetic field of $1.4$ mG and a value $\xi=0.55$ for the surface filling factor of the cloud. The compression factor at the shock, relevant for the purpose of calculating the spectrum of reaccelerated particles is $r_{sh}=4$. 

All spectra show a prominent feature at $\sim 10$ GeV as due to the cutoff induced by particle acceleration at the shock (for He nuclei this corresponds to an energy per nucleon of $\sim 5$ GeV/n). At higher energies, particles only suffer compression both at the shock and in the crushed cloud. One can see that the combined effect of reacceleration and compression is that of shifting the spectra towards higher energies, while conserving the total number of particles in each distribution. At low energy the effect of ionisation leads to a spectral hardening that ends up manifesting itself around few hundred MeV/n. For electrons, in addition to the effects of reacceleration, compression and ionisation losses, one can notice the effect of radiative losses, and more prominently synchrotron losses, that are responsible for the flux suppression at high energies.

\begin{figure*}[!ht]
   \centering
 \subfigure{ \includegraphics[scale=0.8]{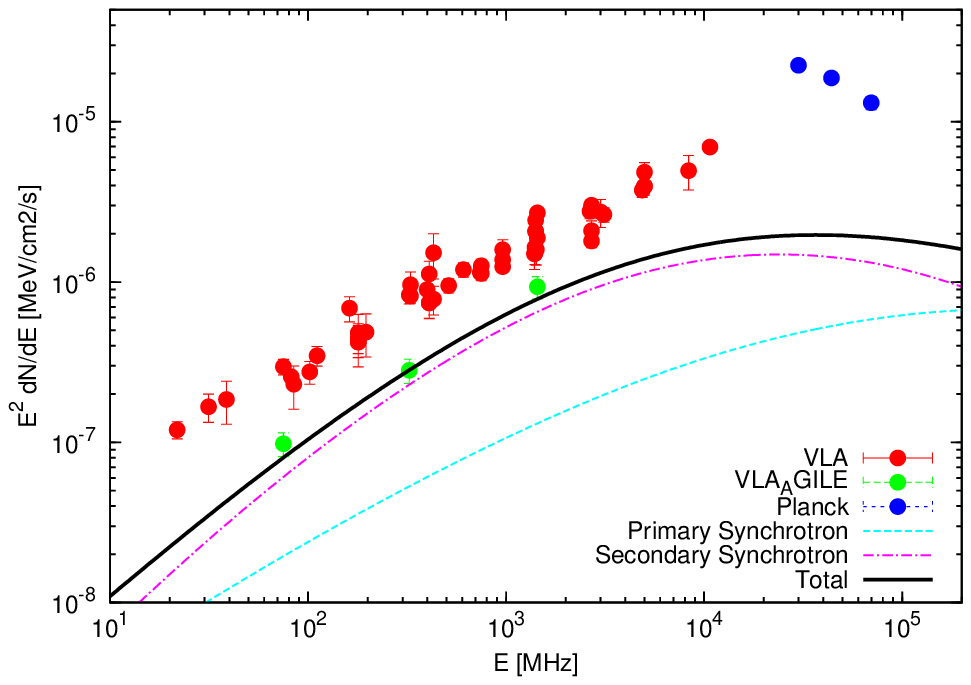} }
 \subfigure{\includegraphics[scale=0.8]{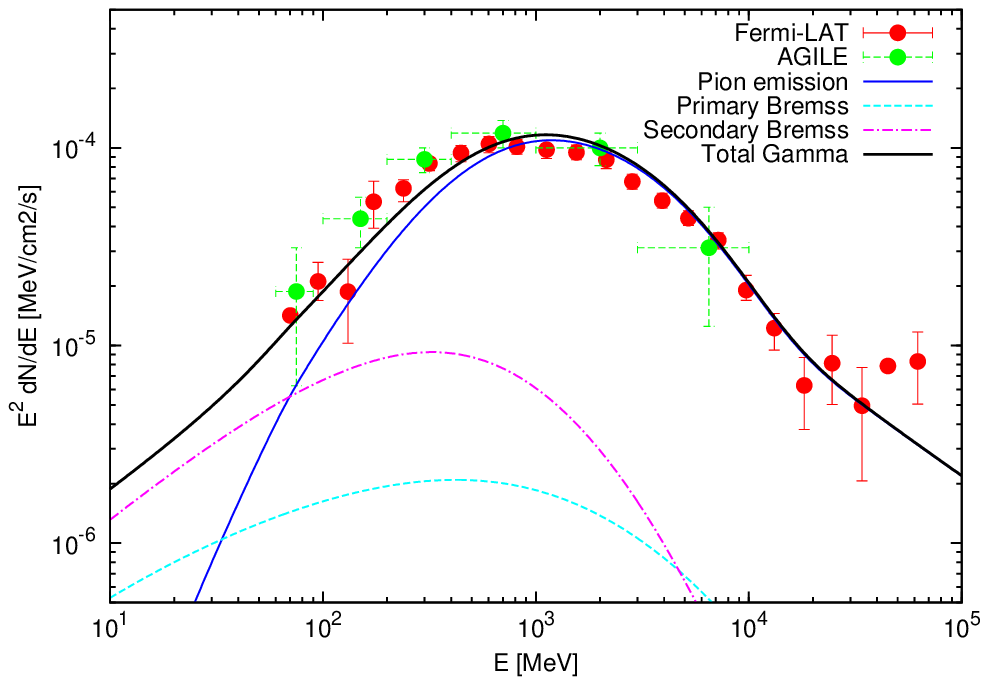} }

   \caption{\textbf{Left}:VLA (red) and Planck (blue) radio data from the whole remnant \citep{castelletti07,planck14} and VLA radio data from the high-energy emitting region (green), plotted together with primary (cyan dashed line), secondary (magenta dot-dashed line) and total (black line) synchrotron radio emission obtained in our best fit reacceleration model. \textbf{Right}: AGILE (green) and Fermi-LAT (red) $\gamma$-ray points \citep{cardillo14_W44,ackermann13_W44} plotted with $\gamma$-ray emission from pion decay (blue dotted line), emission due to bremsstrahlung of primary (cyan dashed line) and secondary (magenta dot-dashed line) electrons, and total emission (black line).}
   \label{Fig:reacceleration}
  \end{figure*}

In Fig.~\ref{Fig:reacceleration}, we show the SED resulting from our best fit reacceleration model, corresponding to the parameters listed above. The left panel shows the radio emission compared with Planck and VLA data from the whole remnant \citep{castelletti07,planck14} and the one from the region where AGILE detected $\gamma$-ray emission \citep{cardillo14_W44}: the total synchrotron emission computed within our best model agrees in slope and normalization with observations. 

For the parameters that best fit the SED of W44 the contribution to radio emission at all observed frequencies is dominated (by a factor of a few) by secondary electrons produced in inelastic $pp$ collisions, as shown in Fig.~\ref{Fig:reacceleration} (left panel). Our prediction is compared with the radio emission from the same region that the AGILE data points in the gamma ray band refer to. 

The conclusion of a dominant secondary contribution is qualitatively similar to the one reached by \cite{uchiyama10_W44}, although with different values of the parameters, but at odds with that of \cite{lee15_W44}, that find instead a prominent contribution of primary electrons (by about two orders of magnitude). The authors suggest that this difference between their results and those of \cite{uchiyama10_W44} may be due to the different account of the time dependence of the problem. It is however difficult to resolve a difference by orders of magnitude in this way, especially taking into account that the values of parameters adopted by both \cite{uchiyama10_W44} and \cite{lee15_W44} are very similar and even the filling factor $f$ that they infer is very similar, $f\sim 0.2-0.5$. 

One point that we wish to emphasise is that the value $f=0.2$, considered by \cite{lee15_W44} as comfortable, being sufficiently less than unity, is in fact rather problematic if one translates it to a filling factor in surface $\xi$: for the same values of parameters as the other authors, we get $\xi>1$.  For the parameters we adopted here, we estimate $\xi=0.55$, less than unity but still suggesting that a large fraction of the SNR surface is involved in the gamma ray emission. 

In the left panel of Fig.~\ref{Fig:reacceleration}, the decrease in the radio emission at frequency above $\sim 10$ GHz is due to the cutoff in the spectrum of reaccelerated protons, that reflects in a suppression in the spectrum of secondary electrons. Notice that radio emission at $\gtrsim 10$ GHz is produced by electrons with energy $\gtrsim 1-2$ GeV, deriving from protons with energy about 10-20 times larger. 

The gamma ray emission from the same region is plotted in the right panel of Fig.~\ref{Fig:reacceleration} and compared with the AGILE \citep{cardillo14_W44} and Fermi-LAT \citep{ackermann13_W44} data points. Most contribution to the gamma ray emission comes from pion decays and in fact the pion bump is very clear. The flux decrease at $E\gtrsim 1$ GeV is directly related to the acceleration cutoff at $\sim 10$ GeV/n in the proton and He spectrum. At energies around and below $\sim 100$ MeV the contribution of the bremsstrahlung emission of secondary electrons is non negligible. Primary electrons play a subdominant role in the production of gamma rays as they do for radio emission. The hardening in the gamma ray emission at $E\gtrsim 10$ GeV is due to the transition to CRs that have been not reaccelerated at the shock (because of lack of enough time to do so) but are nevertheless compressed adiabatically in the shock and crushed cloud region. 

Notice that, contrary to the findings of \cite{uchiyama10_W44} and \cite{lee15_W44} that had to assume the existence of a  spectral break (in addition to an acceleration cutoff) to fit the SED of W44, we obtain a good fit to the data with just the acceleration cutoff. The break in the proton spectrum, already invoked by both AGILE and Fermi-LAT \citep{cardillo14_W44,ackermann13_W44}, is not required by the reacceleration scenario discussed in this section.

One might wonder whether the requirement of a large filling factor derives from the unrealistic assumption that the density of the cloud be uniform, while in reality these objects are expected to be very clumpy and with a high density contrast between the clumps and the interclump medium \citep{bykov00}. We did the attempt to consider the evolution of our SNR in a cloud filled almost totally by inter-clump medium (90$\%$ of the volume) and only for a small fraction (10$\%$ of the volume) by clumps \citep{slane14}. We assumed for the inter-clump medium a density $n_{IC}=20$ cm$^{-3}$ \citep{bykov00} , and knowing that the average density has to be $n_{0}\sim200$ cm$^{-3}$ \citep{yoshiike13_W44}, we derived for the clumps a density $n_{cl}\sim2000$ cm$^{-3}$. We compared the gamma-ray emission resulting from reacceleration of Galactic CRs in the case of a homogeneous medium and in that of a clumpy medium with the same average density. We found that the emission is weaker in the second case, and as a consequence an even larger filling factor is estimated for the cloud. This result, which might seem counterintuitive, is explained in the following. The diffusion coefficient in our model scales with density as: $D(E)\propto n_{0}^{-1/6}$. The dense clumps will have a size that is comparable with the turbulence correlation length $L_{cl}\approx L_c\sim 10^{17}$ cm (see Sec. \ref{Sec:Results}). As a consequence, the time spent by energetic particles in the clumps $t_{cl}$ will be negligible with respect to the time they spend in the interclump medium $t_{cloud}$, being: $t_{cl}\approx\eta (L_{cl}^2)/D_{cl} \approx10^{-4} t_{cloud}$, where $\eta\approx 10\%$ is the volume filling factor of the clumps. This implies that the clumps play no role in the energy evolution of the particles: the latter will simply evolve in a lower density medium and suffer less losses than in the homogeneous case. On the other hand, we expect, and assume in our calculation, that both reacceleration and compression will only take place in the crushed inter-clump medium, which covers now 10\% less of the volume and, especially, has a much lower density than in the homogeneous case . As a result, both the flux of reaccelerated particles and the average post-shock target density for p-p scattering are much lower, and so is the gamma-ray emissivity. Our conclusion is that taking into account inhomogeneity in the simplified way illustrated above leads to estimate even larger filling factors.

 %^^^^^^^^^^^^^^^^^^^^^^^^^^^^^^^^^^^^^^^^^^^^^^^^^^^^^^^^ Fig. hadronic distributions ^^^^^^^^^^^^^^^^^^^^^^^^^^^^^^^^^^
%^^^^^^^^^^^^^^^^^^^^^^^^^^^^^^^^^^^^^^^^^^^^^^^^^^^^^^^^ end Fig. hadronic distributions ^^^^^^^^^^^^^^^^^^^^^^^^^^^^^^^^^^

%-------------------------------

\subsection{Contribution from Acceleration}
\label{sec:AccMod}

The pervasive presence of CRs throughout the Galaxy makes the scenario illustrated above, with reacceleration and compression in the crushed cloud, rather compelling. However, as we pointed out above, the surface filling factor $\xi$ that we obtain is rather close to unity and we consider this to be somewhat disturbing. Hence, we think that there is room for speculation that there may be a contribution to the SED of W44 coming from freshly accelerated particles. In this section we derive a fit to the data in terms of the product $\xi\times\xi_{CR}$ (the spectra are degenerate with respect to this product).

\begin{figure*}[!ht]
   \centering
\subfigure{\includegraphics[scale=0.8]{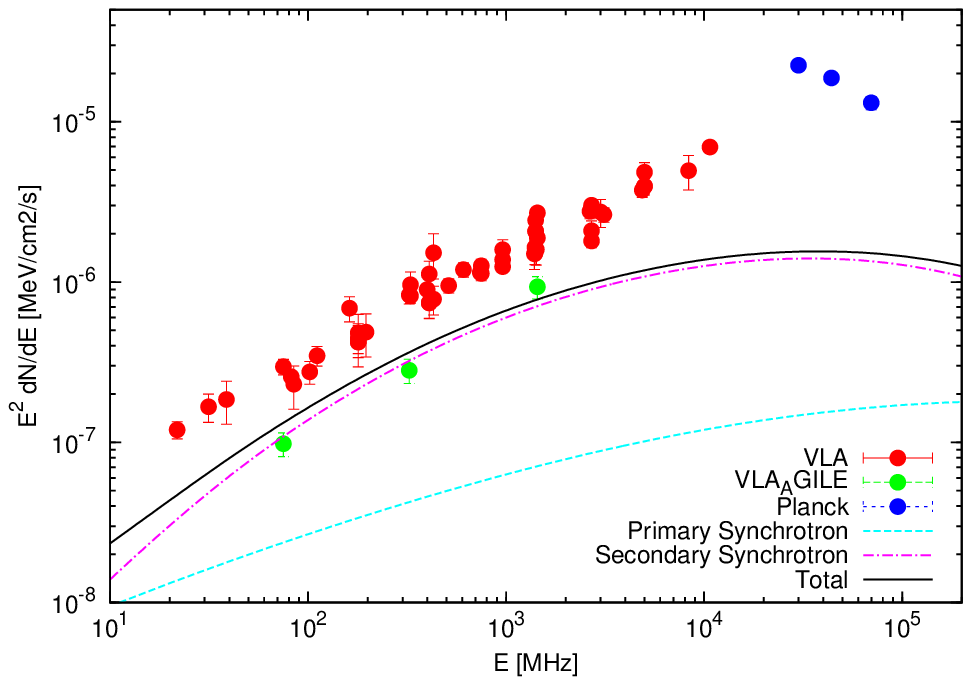} }
 \subfigure{\includegraphics[scale=0.8]{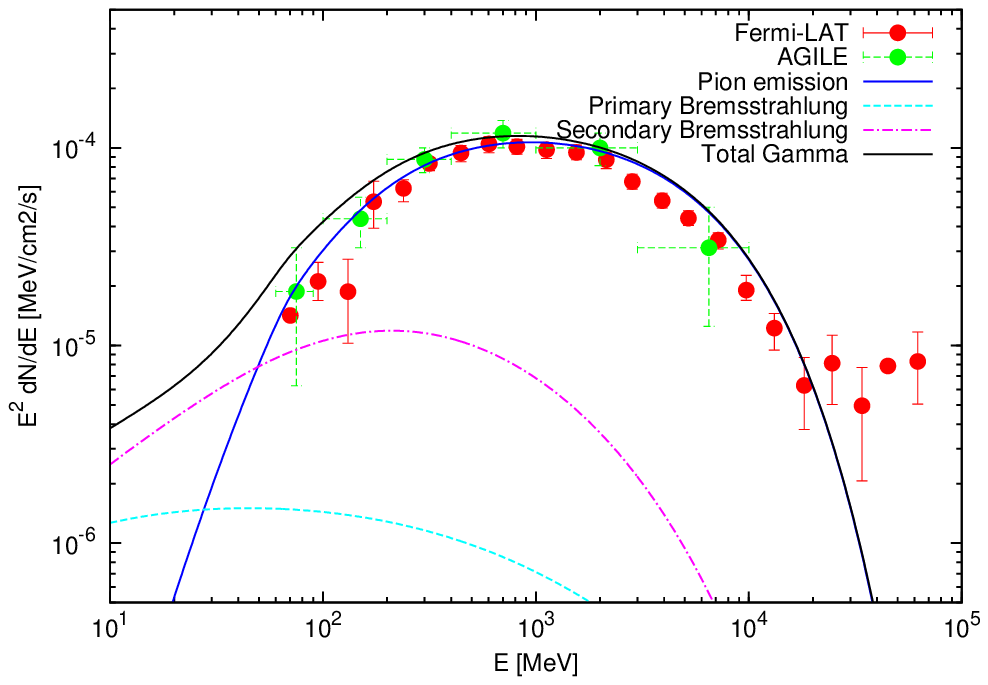} }
   \caption{Radio (left panel) and $\gamma$-ray emission (right panel) from SNR W44 in a model where only freshly accelerated particles are taken into account. The fit is obtained with $\xi\times \xi_{CR}=6.4\times 10^{-5}$. The data points are as in Fig. \ref{Fig:reacceleration}. }
   \label{Fig:acc}
  \end{figure*}

Our results are shown in Fig.~\ref{Fig:acc}, compared with the radio and gamma ray data points. One can see that a qualitatively good fit is achieved, although not as good as in the reacceleration scenario. The curves illustrated in the figure are obtained assuming an interaction time with the cloud of 8400 years and $B_{0}=28.3 \mu G$, a shock velocity $v_{s}=130$ km/s and a parameter combination that leads to $E_{max}=19$ GeV. The normalisation is such that the data are best fit for $\xi \times \xi_{CR}=6.4\times 10^{-5}$. This simple exercise shows that the problem of large values of $\xi$ discussed above for the case of pure reacceleration of Galactic CRs can be alleviated if a small efficiency of fresh CR acceleration is assumed, at the level of $\xi_{CR}\sim 10^{-4}$.

%^^^^^^^^^^^^^^^^^^^^^^^^^^^^^^^^^^^^^^^^^^^^^^^^^^^^^^^^ Fig. hadronic distributions ^^^^^^^^^^^^^^^^^^^^^^^^^^^^^^^^^^
%^^^^^^^^^^^^^^^^^^^^^^^^^^^^^^^^^^^^^^^^^^^^^^^^^^^^^^^^ end Fig. hadronic distributions ^^^^^^^^^^^^^^^^^^^^^^^^^^^^^^^^^^

\subsection{Compression alone}
\label{sec:compression}

As discussed above, even the re-energization of Galactic CRs at the shock of W44 up to a maximum energy of order $\sim 10$ GeV requires relatively strong turbulence to be present at the shock. This condition could be difficult to realise in the case of efficient damping, as would result from the presence of even a small fraction of neutrals in the pre-shock medium \citep{bykov00}. The ion neutral-damping rate is \citep{drury96,ptuskin03} 
\begin{equation}
\Gamma_{\rm IN}=
\left\{
\begin{array}{cc}
\left(\frac{k}{k_c}\right)^2{\nu_{\rm IN}} & k>k_c\\
\nu_{\rm IN} & k<k_c
\end{array}
\right.
\end{equation}
where $\nu_{\rm IN}=8.4\times10^{-9}{\rm s}^{-1}\ \left(T/10^{4}K\right)^{0.4}n_{H}$ is the ion neutral collision frequency, with $T$ the temperature and $n_H$ the density of neutrals in units of ${\rm cm}^{-3}$. The critical wavenumber separating the two regimes is $k_c=(\nu_{\rm IN}/v_A)(n_i/n_H)$, which corresponds to perturbations of wavelength resonant with particles of energy $E_c\approx 2.5$ GeV in the regime of density and magnetic field strength we are considering. Above this energy the damping becomes progressively inefficient, so that the meaningful timescales to compare in order to assess whether efficient acceleration (or reacceleration) is possible, are the timescale for ion neutral damping of turbulence at wavelength $k_c$ and the time $t_{\rm int}$ for which the shock has been interacting with the cloud. We find that $\Gamma_{\rm IN}t_{\rm int}<1$ for $n_H<5 \times 10^{-4}\ {\rm cm}^{-3}$, a rather stringent requirement.

The latter condition on the density of neutrals is the appropriate one to ensure that pre-existing turbulence is not damped at the scales that are relevant for particle acceleration in the source. The reaccelerated particles provide themselves a potential source of turbulence, however, whose growth is possible if $\Gamma_{\rm CR}>\Gamma_{\rm IN}$ with $\Gamma_{\rm CR}$ the growth rate of the resonant streaming instability associated with reaccelerated particles. We computed $\Gamma_{\rm CR}$ based on the spectrum of reaccelerated particles that is required to fit the data and found that in our case turbulence can efficiently grow if $n_H< 5 \times 10^{-3} {\rm cm}^{-3}$. While more relaxed than in the former case, this condition on the level of ionization of the medium still appears rather stringent, making it especially meaningful to speculate on the possibility that the level of turbulence is low enough that, in the energy range that is relevant for the detected emission, Galactic CRs are only compressed at the shock and in the crushed cloud, without substantial diffusive shock reacceleration.

\begin{figure*}[!ht]
   \centering
\subfigure{\includegraphics[scale=0.8]{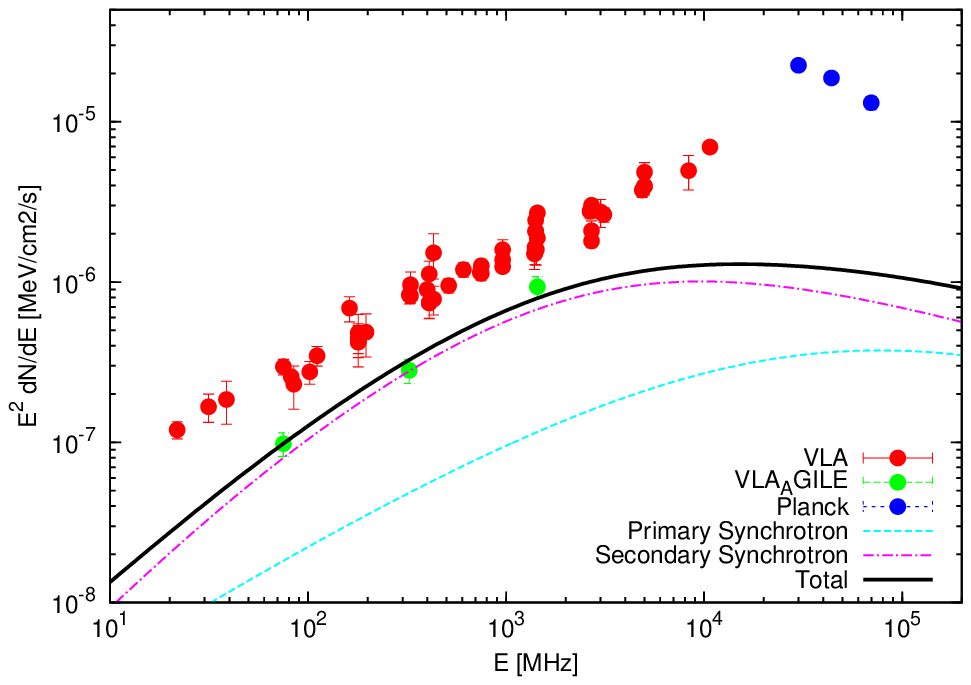} }
 \subfigure{\includegraphics[scale=0.8]{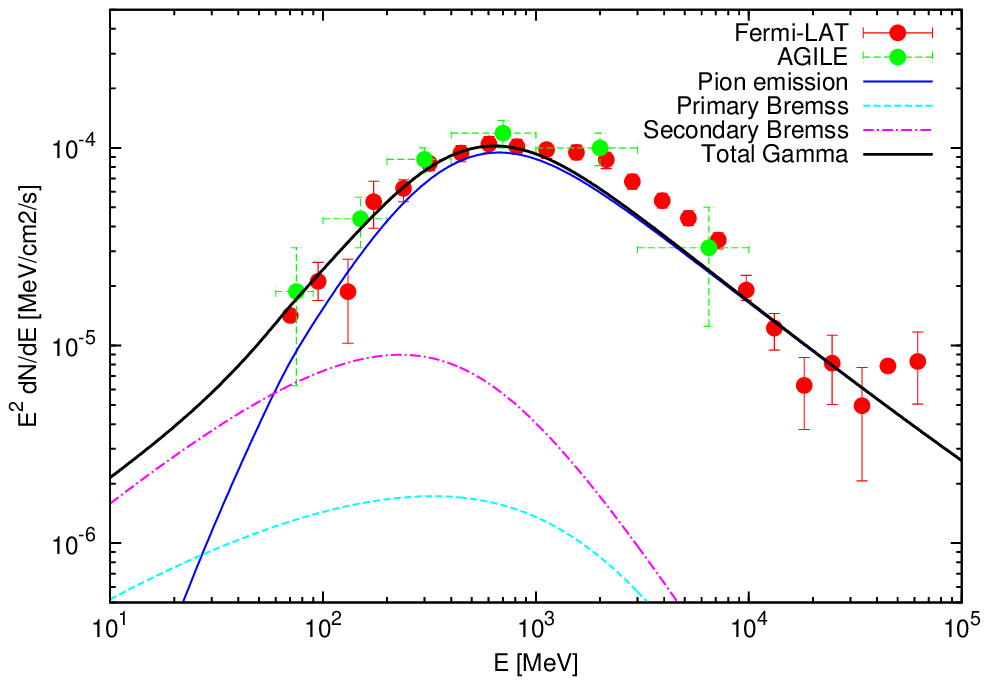} }
   \caption{Radio and $\gamma$-ray emission from SNR W44 assuming that Galactic CRs are only compressed at the shock and in the crushed cloud. In the left (right) panel we show the radio (gamma ray) emission. Data are as in Fig. \ref{Fig:reacceleration}.}
   \label{Fig:compression}
  \end{figure*}

The SED obtained in the compression-only scenario is shown in Fig. \ref{Fig:compression}, in the radio band (left panel) and in the gamma ray band (right panel). The values of the parameters are the same adopted for the reacceleration model, but the surface filling factor is $\xi=0.65$. The main effect of ignoring reacceleration is that the bump in the gamma ray spectrum in the region $1-10$ GeV disappears, so that the fit becomes somewhat worse. On the other hand, the radio emission, being related to low energy secondary electrons, can be well explained even assuming that Galactic CRs are simply compressed. 

%%%%%%%%%%%%%%%%%%%%%%%%%%%%%%%%%%%%%%%
\section{Discussion and conclusions}
\label{sec:Conclusions}
%%%%%%%%%%%%%%%%%%%%%%%%%%%%%%%%%%%%%%%
After the detection of low-energy $\gamma$-ray emission from SNR~W44 by AGILE \citep{giuliani11_W44} and Fermi-LAT \citep{ackermann13_W44}, a number of different models have been put forward to explain observations. The detection was initially presented as the first unequivocal evidence of CR acceleration in a SNR, although the old age of the remnant  \citep{giuliani11_W44, abdo10_W44, ackermann13_W44, cardillo14_W44} and its low shock velocity do not make W44 a likely location for efficient CR acceleration. On the other hand, part of the shell of SNR W44 is clearly interacting with a dense molecular cloud, which would imply that a dense target for hadronic interactions is present. \cite{uchiyama10_W44} and later  \cite{lee15_W44} discussed the possibility that the SED of W44 could be explained in terms of diffusive shock reacceleration of Galactic CRs and further compression in the crushed cloud that the SNR is impacting upon. In these conditions the shock slows down to $v_s\approx 100-150$ km$/$s and a thin radiative shell with density $10^3-10^4$ cm$^{-3}$ is formed, where magnetic field strength is also enhanced. 

The calculations of  \cite{uchiyama10_W44} and \cite{lee15_W44} have several points in common but also show several important differences: they adopt the same parametrisation for the low energy behavior of the Galactic CR spectrum, assumed to be made of protons and electrons only. In both cases the fit to the SED requires a spectral break at $\sim 10$ GeV, attributed to a mechanism previously discussed by \cite{malkov11_W44}, in addition to a high energy cutoff induced by reacceleration. In both cases a volume filling factor of $0.2-0.5$ is assumed to fit the data. However, the calculations of \cite{uchiyama10_W44} suggest that most emission in the radio band is due to synchrotron emission of secondary electrons, produced in inelastic pp scattering, while the primary contribution dominates upon the secondary one by two orders of magnitude in the calculations of \cite{lee15_W44}. The authors mention that this discrepancy might be due to the different way the temporal evolution is treated: while  \cite{lee15_W44} follow the evolution of the shock in time, \cite{uchiyama10_W44} assume a simple model in which the shock moves with constant velocity inside the cloud. The very recent work of \cite{tang15} leads to a conclusion very similar to ours: middle-aged SNR spectra can be explained by simple reacceleration plus compression of galactic CRs, without invoking any HE spectral break. However, this conclusion is reached based on a very different model, that does not take into account particle energy losses, which on the contrary we find to be important, and does not include leptonic emission, which we also find to be important, not only in the low frequency range, but also in the gamma-ray band.

In the present paper we carried out an analysis similar to that by \cite{uchiyama10_W44}, but with some relevant improvements, that we list hereafter. 1) Instead of parametrising the Galactic CR spectrum in the ISM at low energies, we adopted the spectra as measured by Voyager 1 and analysed by \cite{potgieter14}. 2) Our calculations account for protons and He nuclei in Galactic CRs, as well as primary electrons (the latter are also measured by Voyager I, although at a time earlier than reaching the heliopause, so that some correction for residual solar modulation was needed, as discussed by  \cite{potgieter14}). 3) We investigated the possibility that the reacceleration occurred at the SN shock in its most basic form, without invoking the spectral break at $10$ GeV, as was done by \cite{uchiyama10_W44} and \cite{lee15_W44}. 4) We discuss the implications of a situation in which, perhaps due to lack of turbulence in the cloud at the relevant scales, reacceleration was inefficient so that Galactic CRs are only compressed in the crushed cloud. 5) We reformulated the problem in terms of a surface filling factor, having in mind that the radiative shell is thin and unlikely to fill a large fraction of the SNR volume ($f\sim 0.2-0.5$ was found in previous calculations from fits to the data).

We confirm that the reacceleration and compression of Galactic CRs can explain the SED of W44, although using the same parameters as \cite{uchiyama10_W44} we find that $f\sim 0.5$ corresponds to $\xi>1$, which is clearly unphysical. We obtain a satisfactory description of the radio and gamma ray emission from W44 by assuming a somewhat larger shock velocity $v_{s}=130$ km/s and a magnetic field in the cloud $B_{0}=34\mu$G. The shock is assumed to have been interacting with the cloud of average density $200~\rm cm^{-3}$ for the last $8400$ years. This corresponds to $\xi=0.55$, namely more than half of the surface of the shell involved in the interaction with the dense cloud. Although physically allowed, even such value of $\xi$ appears rather large, though not ruled out. We tried to evaluate whether taking into account the likely clumpiness of the cloud could help alleviate this requirement, but found that in the simplest picture, in which clumps occupy 10\% of the volume, have an average size of the order of the estimated coherence length of the turbulence and a density contrast of about 100, the required filling factor of the cloud becomes even larger.

% $L_{cl}\sim L_{c}\sim(L_{cloud}/25)$ : (1) the dimension of the cloud in our model is $L_{c}\sim v_{sh}t_{int}\sim 2.4\times10^{18}$;(2) since a Kolmogorov turbulence spectrum has almost all its power at large scales, on average we can assume all the turbulence focused at the correlation length scale, $L_{c}\sim10^{17}$, we take the dimension of a clump, $L_{cl}\sim L_{c}$, in order to contain all turbulence scales;(3) the initial magnetic field can be expressed as $B_{0}\propto\sqrt{n_{0}}$, we find that the time spent by particles in the volume filled with dense clumps is $t_{cl}\approx\eta (L_{cl}^2)/D_{cl} \approx10^{-4} t_{cloud}\ll {t_{cloud}}$ where $\eta$ is the volume filling factor filled with clumps, $\eta\sim0.1$.}

The scenario in which the SED of W44 is interpreted as the result of reacceleration and compression of Galactic CRs has many intriguing aspects: the hard radio spectrum finds a natural explanation in terms of low energy secondary electrons. The same hadronic interactions that produce secondary electrons also lead to pion production and decay that fits the gamma ray data. It is important to realise that no free parameter is involved in the reacceleration of Galactic CRs, except for the maximum particle energy, namely no CR acceleration efficiency needs to be specified, because all relevant particles are already above threshold for injection. Most emission is explained in terms of low energy Galactic CRs (with very hard spectrum, as measured by Voyager I), eventually re-energized at the shock and further compressed in the crushed cloud. The shape of the gamma ray spectrum is explained in terms of a cutoff of the re-energized protons and electrons, with maximum energy $E_{max}\sim 10$ GeV, derived by assuming a Kolmogorov turbulence in the shock region with $\delta B/B\sim 1$ on a scale $L_{c}\sim 0.1$ pc. Notice that particles with $E>E_{max}$ are not reaccelerated but are still compressed: this effect allows us to achieve a good fit to the gamma ray data even at energies of several tens of GeV (from Fermi-LAT). With the combination of parameters adopted here, the radio spectrum is dominated by synchrotron emission of secondary electrons, in qualitative agreement with the conclusion of \cite{uchiyama10_W44} but at odds with the calculations of \cite{lee15_W44} that lead to a strong dominance of the contribution from primary electrons. 

If the turbulence is weaker in the shock region, then $E_{max}$ can be even smaller than $\sim 10$ GeV. We considered a speculative case in which there is no reacceleration at all but only compression in the crushed cloud. The radio and gamma ray data can be fit even in this case although the quality of the fit is much worse and an even larger value of the surface filling factor is needed, $\xi=0.65$. 

The large values of $\xi$ required by the reacceleration model inspired us in searching for meaningful information about acceleration of fresh CRs at the shock in W44. Assuming again that no reacceleration of Galactic CRs takes place, one can see that a satisfactory fit to the radio and gamma ray emission of W44 can be achieved by assuming the same parameters that describe the best-fitting reacceleration model, except for the unperturbed magnetic field and the maximum particle energy, that now read, respectively, $B_{0}=28$ $\mu$G and $E_{max}=19$ GeV. The data are satisfactorily well fit for $\xi\times\xi_{CR}\simeq 6.4 \times 10^{-5}$, although the quality of the fit is not excellent. The conclusion we can draw from this exercise is that freshly accelerated particles help alleviate the problem of large values of $\xi$. In fact a mild mix of reacceleration and particle acceleration with $\xi_{CR}\sim 10^{-4}$ appears to provide a good fit to the data and lower the required fraction $\xi$ of the shock surface involved in the interaction with the cloud.

 \begin{acknowledgements}
We are very grateful to Mirko Boezio and Marius Potgieter for providing the Voyager data for protons, He nuclei and electrons, Gabriela Castelletti and Gloria Dubner for providing us with the radio data points from the region coincident with the AGILE gamma ray detection. This work was partially funded through Grant PRIN-INAF 2012.
 \end{acknowledgements}

\begin{appendix}
\section{Particles' energy losses}

\label{Sec:app1}
%------------------------------
Nuclear collisions are described as:
%**************
\begin{equation}
-\left(\frac{dE_{p}}{dt}\right)_{pp}=\frac{E_{p}}{\tau_{pp}}=E_{p}f_{in}n_{m}v_{p}(E_{p})\sigma_{pp}(E_{p}^{tot})
\label{eq:pp}
\end{equation}
%***********************
where $E_{p}$ is proton energy, $f_{in}$, taken to be constant, accounts for the inelasticity of the collisions, $v_{p}$ is the particle velocity and $\sigma_{pp}$ is the cross section of the process, which we describe according to \cite{kelner06}.

Ionisation losses for protons are described according to \cite{ginzburg69}, assuming that the plasma is completely ionized: %\begin{displaymath}
\begin{equation}
\begin{split}
&-\left(\frac{dE_{p}}{dt}\right)_{ion,p}=7.62\times 10^{-9}Z^{2}n_{m}\times \\
&\times\left\{ \begin{array}{lll}
\sqrt{\left(\frac{2E_{p}^{0}}{E_{p}}\right)}\left(38.7+\ln\left(\frac{E_{p}}{E_{p}^{0}}\right)-\frac{1}{2}\ln(n_{m})\right)& \textrm{if}\, E_{p}\ll E_{p}^{0}\\
\left(74.1+2\ln\left(\frac{E_{p}^{tot}}{E_{p}^{0}}\right)-\ln(n_{m})\right)& \textrm{if}\, E_{p}^{0}\ll E_{p}\ll \left(\frac{m_{p}}{m_{e}}\right)E_{0}^{p}\\
\left(74.1+\ln\left(\frac{E_{p}^{tot}}{E_{p}^{0}}\right)-\ln(n_{m})\right)& \textrm{if}\, E_{p}\gg \left(\frac{m_{p}}{m_{e}}\right)E_{0}^{p}
\end{array} \right.
\label{eq:p_ion}
\end{split}
\end{equation}
%\end{displaymath}
where the energy lost is expressed in $eV/s$.

Again following \cite{ginzburg69}, ionization losses for ultrarelativistic electrons in a completely ionized medium are described as:
%**************
\begin{equation}\begin{split}
-\left(\frac{dE_{e}}{dt}\right)_{ion,e}&=\frac{2\pi e^{4}n_{m}}{m_{e}c}\left[\ln\left(\frac {E_{e}(m_{e}c)^{2}}{4\pi e^2n_{m}\hbar^2}\right)-\frac{3}{4}\right]=\\
&=7.62\times10^{-9}n_{m}\left[\ln\left(\frac{E_{e}}{E_{e}^{0}}\right)-\ln(n_{m})+73.4\right]
\label{eq:e_ion}
\end{split}
\end{equation} 
in $eV/s$. 

Electrons are also affected by synchrotron losses
%***************
\begin{equation}
-\left(\frac{dE_{e}}{dt}\right)_{syn}=\frac{4}{9}\frac{e^{4}B_{m}^{2}}{m_e^{2}c^{3}}\left(\frac{E_{e}}{E_{e}^{0}}\right)^2\ ,
\label{eq:sync_loss}
\end{equation}
%*****************
inverse Compton Scattering (ICS),
%***************
\begin{equation}
-\left(\frac{dE_{e}}{dt}\right)_{IC}=\frac{32}{9}\pi\left(\frac{e^{2}}{m_ec^{2}}\right)^2cU_{rad}\left(\frac{E_{e}}{E_{e}^{0}}\right)^2\ ,
\label{eq:ic_loss}
\end{equation}
%************
and Bremsstrahlung \citep{ginzburg69}
%*****************
\begin{equation}
\begin{split}
-\left(\frac{dE_{e}}{dt}\right)_{Brem,1}=\frac{4e^{6}n_{m}Z(Z+1)}{m_{e}^{2}c^{4}\hbar}E_{e}\left[\ln\left(\frac{2E_{e}}{E_{e}^{0}}\right)-\frac{1}{3}\right]=\\
=2.2\times10^{-19}{\rm erg\ s^{-1}}\left[\ln\left(\frac{E_{e}}{E_{e}^{0}}\right)+0.36\right]\left(\frac{n_{m}}{1\,cm^{-3}}\right)\left(\frac{E_{e}}{1\,GeV}\right)\ .
\end{split}
\label{eq:Brems_loss_low}
\end{equation}
all expressed in $erg/s$.
%*********************

\section{Radio and $\gamma$-ray emission}
\label{sec:emission}

\subsection{Secondary emission}
In order to compute the $\gamma$-ray emission and secondary pair production, we used the formalism described by \cite{kelner06}. For kinetic particle energies per nucleon $E_{n}=E/n>0.1$ TeV, the number of secondaries in units of $1/(erg/n)/cm^{3}/s$, is:
%****************
\begin{equation}
\Phi_{s,high}(E_{s})=v\,n_{0}\int^{1}_{0}\sigma_{inel}(E_s/x)f(E_s/x)F_s(x,E_s/x)\frac{dx}{x} 
\label{Eq:sec_high}
\end{equation} 
%****************
where $s$ refers to the secondary specie being considered, either electrons or $\gamma$-rays, $x=E_s/E_{n}$, $F_s$ is an analytical function describing the spectral distribution of the secondaries (see \cite{kelner06} for details), $v$ is the nucleon velocity and $f(E_s/x)=f(E_n)$ is the final distribution of primary nucleons of given energy $E_{n}$. We write:
\begin{equation}
f(E_n)=f_{p}(E_{n})+4f_{He}(E_n)
\end{equation}
where $f_{p}(E_{n})$ is the distribution of protons, and $f_{He}$ that of He nuclei.
The inelastic cross section, $\sigma_{inel}$, is:
%****************
\begin{equation}
\sigma_{inel}=(34.3+1.88L+0.25L^{2})\left[1-\left(\frac{E_{th}}{E_{p}}\right)^{4}\right]^{2}\,\,\,\,\,mb
\label{Eq:cross_section}
\end{equation}
%****************
where $L=\ln\left(\frac{E_{p}}{1TeV}\right)$ and  $E_{th}=E_{0,p}+2E_{0,\pi}+E_{0,\pi}^{2}/E_{0,p}$, is the threshold energy for $\pi^{0}$ ($\pi^{\pm}$) production used in order to compute the photon (secondary electrons) spectrum.\\
At lower energies, we used the $\delta$-approximation for secondary emission rate:
%****************
\begin{equation}
\Phi_{s,low}(E_s)=2\times\int^{\infty}_{E_{min}}\frac{f_{\pi}(E_{\pi})}{\sqrt{E_{\pi}^{2}-(m_{\pi}c^2)^{2}}}dE_{\pi}
\label{Eq:sec_low}
\end{equation}
%****************
where the factor 2 accounts for the production of two photons from every neutral pions and both $e^+$ and $e^-$ in the case of charged pions,  $E_{min}=E_s+(m_{\pi}c^{2})^{2}/4E_s$ is the minimum energy the pion must have in order to produce a secondary with energy $E_s$ and $f_{\pi}$ is the production rate of pions with energy $E_\pi$:
%****************
\begin{equation}
f_{\pi}(E_{\pi})=\frac{\tilde{n}}{K_{\pi}}\,v\,n\,\sigma_{inel}\left(m_p c^2+\frac{E_{\pi}}{K_{\pi}}\right)f_{p}\left(\frac{E_{\pi}}{K_{\pi}}\right)
\label{Eq:pions}
\end{equation}
%****************
with  $\tilde{n}$ the number of produced pions for a given proton distribution function and $K_{\pi}=k/\tilde{n}$, where $k$ is the fraction of kinetic energy per nucleon transferred to the pion. The particle spectrum $f_{p}\left(\frac{E_{\pi}}{K_{\pi}}\right)$ is the same used for the high-energy regime and $\frac{E_{\pi}}{K_{\pi}}=E_n$. Following \cite{kelner06}, we fixed $K_{\pi}=0.17$ and found the value of $\tilde{n}$ that leads the continuity of the secondary spectrum at $E=0.1$ TeV . 
Eqs.~\ref{Eq:sec_high} and \ref{Eq:sec_low} provide both the $\gamma$-ray photon production rate and the source term that must replace expression \ref{eq:injection_rate} when the kinetic equation (Eq.~\ref{Eq:kinetic}) is written for secondaries.

\subsection{Synchrotron emission}
The synchrotron emission rate, in units of $1/cm^{3}/s$, is:
%****************
\begin{equation}
\Phi_{syn}(E_{\gamma})=\frac{\sqrt{3}e^{3}B_{m}}{4\pi E_{e}^{0}h}\int f_{e}(E_{e})dE_{e}\,R(\omega/\omega_{c})
\label{Eq:synch}
\end{equation}
%****************
where $R(\omega/\omega_{c})$ is the function describing synchrotron radiation by a single electron in a magnetic field with chaotic directions and $\omega_{c}=\frac{1.5eB_{m}p^{2}}{m_{e}^{3}c^{3}}$ is the characteristic synchrotron frequency \citep{zirakashvili07}.  

\subsection{Bremsstrahlung}
The $\gamma$-ray emissivity due to bremsstrahlung is described as:
%****************
\begin{equation}
\Phi_{Brem}(E_{\gamma})=1.8\,c\,n\int dE_{e}f_{e}(E_{e})\frac{d\sigma}{dE_{\gamma}}
\label{Eq:Brems}
\end{equation}
%****************
in units of ${\rm erg}^{-1} {\rm cm}^{-3}\ {\rm s}^{-1}$ \citep{blumenthal70}, where we used the Bremsstrahlung differential cross section given by \cite{ginzburg69}:
%****************
\begin{equation}
\frac{d\sigma}{dE_{\gamma}}=4\alpha_{fs} Z^{2}r_{0}^{2}\frac{dE_{\gamma}}{E_{\gamma}}\left[\left(1+\left(1-\frac{E_{\gamma}}{E_{e}}\right)^{2}\right)\phi_{1}+\left(1-\frac{E_{\gamma}}{E_{e}}\right)\phi_{2}\right]
\label{Eq:Brems_cross_section}
\end{equation}
%****************
where $\alpha_{fs}=\frac{e^{2}}{\hbar c}$ is the fine structure constant, $r_{0}=\frac{e^{2}}{m_{e}c^{2}}$ is electron classical radius and $\phi_{1,2}$ are functions that depend on the electron and photon energies, and are different for the case of strong and weak shielding (our case) \citep{ginzburg69}. The factor $1.8$ in Eq~\ref{Eq:Brems} takes into account the presence of different kinds of nuclei. 

\subsection{Inverse Compton Scattering}
One last process that contributes to $\gamma$-ray emission is Inverse Compton scattering, for which we write the emissivity as:
%****************
\begin{equation}\begin{split}
\Phi_{IC}(E_{\gamma})&=\frac{2\pi r_{0}^{2}c}{\gamma^{2}}\int\int\frac{f_{ph}(E_{ph})dE_{ph}}{E}\times\\
&\times\left[2\,q\,\ln{q}+(1+2q)(1-q)+\frac{1}{2}\frac{\left(\Gamma_{e}q\right)^{2}}{1+\Gamma_{e}q}(1-q)\right]
\label{Eq:IC}
\end{split}
\end{equation}
%****************
where $\Gamma_{E_{ph}}=4E_{ph}\gamma/mc^{2}$, $q=E_{e}/\Gamma_{E_{ph}}(1-E_{e})$, $f_{ph}(E_{ph})$ is the distribution of target photons, $E_{ph}$ is the target photon energy and $E_{e}$ is the electron energy. However, we will show that, for the parameter s values appropriate to describe SNR~W44, the IC contribution to the $\gamma$-ray emission is negligible.

\end{appendix}

\end{document}